# Forbidden zones for circular regular orbits of the Moons in Solar system, R3BP

**Sergey V. Ershkov**,

Sternberg Astronomical Institute,

M.V. Lomonosov's Moscow State University,

13 Universitetskij prospect, Moscow 119992, Russia

e-mail: sergej-ershkov@yandex.ru

**Abstract:** Previously, we have considered the equations of motion of the three-body problem in a *Lagrange form* (which means a consideration of relative motions of 3-bodies in regard to each other). Analyzing such a system of equations, we considered the case of small-body motion of negligible mass $m_3$ around the 2-nd of two giant-bodies $m_1$, $m_2$ (*which are rotating around their common centre of masses on Kepler's trajectories*), the mass of which is assumed to be less than the mass of central body.

In the current development, we have derived a key parameter η that determines the character of *quasi-circular* motion of the small 3-rd body $m_3$ relative to the 2-nd body $m_2$ (Planet). Namely, by making several approximations in the equations of motion of the three-body problem, such the system could be reduced to the key governing *Riccati*-type ordinary differential equations.

Under assumptions of R3BP (restricted three-body problem), we additionally note that *Riccati*-type ODEs above should have the invariant form if the key governing (dimensionless) parameter η remains in the range $10^{-2} \div 10^{-3}$. Such an amazing fact let us evaluate the *forbidden zones* for Moons orbits in the inner Solar system or the zones of the meanings of distances (*between Moon-Planet*) for which the motion of small body could be predicted to be *unstable* according to basic features of the solutions of *Riccati*-type.

# Introduction.

Predictable stability of the motion of the small bodies in Solar system is the ancient problem in the field of celestial mechanics which leading scientists have been trying to solve during last 300 years. An elegant ansatz to present such a problem from a point of view of relative motions in restricted three-body problem (R3BP) is proposed here.

In our previous article [1], we have explored the stability of the Moons motion in Solar system, which are supposed to be rotating around their Planets on *quasi-elliptic* orbits (such Moons are massive enough to compose their own central field of gravitation or which are massive enough to have achieved hydrostatic equilibrium).

The main aim of the current development is definetely to evaluate the sizes of *forbidden* zones around the Planets of inner Solar system, i.e. the zones of the predicted (natural) instability of the motion of the small-bodies such as moons of the Planets, or asteroids, or the spacecraft, drifting on the path around the Planet.

We should especially note the appropriate restrictions to the derivation: the current results are correct only in terms of *restricted* three-body problem [2-5], but we have not taken into consideration the effects of *photogravitational* restricted three-body problem [6-7] or the relativistic effects as well as the effect of variable masses of 3-bodies [8]. We also neglect the additional influence of *Yarkovsky* effect of non-gravitational nature [9].

To avoid ambiguity, we should consider a *relative* motion in three-body problem [3].

# 1. Equations of motion.

Let us consider the system of ODE for *relative* motion of restricted three-body problem in barycentric Cartesian co-ordinate system, at given initial conditions, see Eq. (5) in [1] (where $\gamma$ is the gravitational constant).

Under assumption of the *restricted* three-body problem, we assume that the mass of small 3-rd body $m_3 \ll m_1, m_2$, accordingly; besides, $\boldsymbol{R}_{2,3}$ means the radius-vector of relative motion between bodies $m_2, m_3$, whereas $\boldsymbol{R}_{1,2}$ means the radius-vector of relative motion between bodies $m_1, m_2$, accordingly.

According to the assumptions of the *restricted* three-body problem, let us consider the case of *quasi-periodic* motion of a small 3-rd body $m_3$ as a moon around the 2-nd body $m_2$ under additional assumption $|\boldsymbol{R}_{2,3}| \ll |\boldsymbol{R}_{1,2}|$. Also we should note that vector $\boldsymbol{R}_{1,2}$ is assumed to be corresponding to the case of strictly *periodic* relative motion of 2 massive bodies $m_1, m_2$ (*which are rotating around their common centre of masses on Kepler's trajectories*).

In such a case, it was proved [1] that equations of R3BP could be reduced for the *restricted* mutual motions of bodies $m_2, m_3$ (as first approximation) as below:

$$\boldsymbol{R}_{2,3}'' + \gamma(m_2+m_3)\frac{\boldsymbol{R}_{2,3}}{|\boldsymbol{R}_{2,3}|^3} + \frac{\gamma m_1}{|\boldsymbol{R}_{1,2}|^3}\left(1+3\cos\alpha\frac{|\boldsymbol{R}_{2,3}|}{|\boldsymbol{R}_{1,2}|}\right)\boldsymbol{R}_{2,3} \cong -3\cos\alpha\left(\frac{\gamma m_1}{|\boldsymbol{R}_{1,2}|^3}\boldsymbol{R}_{1,2}\right)\frac{|\boldsymbol{R}_{2,3}|}{|\boldsymbol{R}_{1,2}|}, \quad (1.1)$$

- here $\alpha$ is the angle between the radius-vectors $\boldsymbol{R}_{2,3}$ and $\boldsymbol{R}_{1,2}$.

Obviously, the assumption $|\boldsymbol{R}_{2,3}| \ll |\boldsymbol{R}_{1,2}|$ yields from Eq. (1.1) as below

$$\boldsymbol{R}_{2,3}'' + \gamma(m_2+m_3)\frac{\boldsymbol{R}_{2,3}}{|\boldsymbol{R}_{2,3}|^3} + \frac{\gamma m_1}{|\boldsymbol{R}_{1,2}|^3}\boldsymbol{R}_{2,3} \cong -3\cos\alpha\left(\frac{\gamma m_1}{|\boldsymbol{R}_{1,2}|^3}\boldsymbol{R}_{1,2}\right)\frac{|\boldsymbol{R}_{2,3}|}{|\boldsymbol{R}_{1,2}|}, \quad (1.1')$$

Let us consider the case of *quasi-circular* motion of the small 3-rd body $m_3$ as a moon around the 2-nd body $m_2$. It means that $(|\boldsymbol{R}_{2,3}|/|\boldsymbol{R}_{1,2}|) = \varepsilon \cong \text{const} \to 0$ for a sufficiently long time-period $\tau$. But we should consider such a time-period $\tau$ to be *much less* than the period $T$ of evolving the body $m_2$ around the common centre of masses bodies $m_1$, $m_2$ on Kepler's trajectories, $\tau \ll T$. Thus, the last equation (1.1′) could be transformed for the periodic changing of the angle $\alpha$ above as below

$$\int_0^{\pi/\omega} \left( \boldsymbol{R}_{2,3}'' + \gamma(m_2+m_3)\frac{\boldsymbol{R}_{2,3}}{|\boldsymbol{R}_{2,3}|^3} + \frac{\gamma m_1}{|\boldsymbol{R}_{1,2}|^3}\boldsymbol{R}_{2,3} \right) dt \cong -3\varepsilon \cdot \left( \frac{\gamma m_1}{|\boldsymbol{R}_{1,2}|^3}\boldsymbol{R}_{1,2} \right) \cdot \int_0^{\pi/\omega} (\cos\alpha)\,dt \, ,$$

$$\int_0^{\pi/\omega} (\cos\omega t)\,dt \cong 0 \quad (\omega \cong const) \quad \Rightarrow \quad \int_0^{\pi/\omega} \left( \boldsymbol{R}_{2,3}'' + \left( \frac{\gamma(m_2+m_3)}{|\boldsymbol{R}_{2,3}|^3} + \frac{\gamma m_1}{|\boldsymbol{R}_{1,2}|^3} \right) \cdot \boldsymbol{R}_{2,3} \right) dt \cong 0 \quad (1.2)$$

In [1] Eq. (1.2) has been presented in the form below

$$\boldsymbol{R}_{2,3}'' + \gamma(1+\xi+\eta)\cdot m_2 \cdot \frac{\boldsymbol{R}_{2,3}}{|\boldsymbol{R}_{2,3}|^3} = 0 , \qquad (1.3)$$

$$\xi = \left( \frac{m_1}{m_2} \cdot \frac{|\boldsymbol{R}_{2,3}|^3}{|\boldsymbol{R}_{1,2}|^3} \right), \quad \eta = \left( \frac{m_3}{m_2} \right)$$

- where Eq. (1.3) describes the relative motion of the centre of mass of 2-nd giant-body $m_2$ (Planet) and the centre of mass of 3-rd body (Moon) with the effective mass ($\xi \cdot m_2 + m_3$), which are rotating around their common centre of masses on the stable Kepler's elliptic trajectories. Besides, if the dimensionless parameters $\xi, \eta \to 0$ then equation (1.2) should describe a *quasi-circle* motion of 3-rd body (Moon) around the 2-nd body $m_2$ (Planet), see Tab.1 in [1].

But we could also present Eq. (1.1) in the form below:

$$\boldsymbol{R}_{2,3}'' + (1+\zeta) \cdot \left( \frac{\gamma m_1}{|\boldsymbol{R}_{1,2}|^3} \right) \cdot \boldsymbol{R}_{2,3} = 0 , \qquad (1.4)$$

$$\zeta = \left( \frac{(m_2+m_3)}{m_1} \cdot \frac{|\boldsymbol{R}_{1,2}|^3}{|\boldsymbol{R}_{2,3}|^3} \right),$$

- where Eq. (1.4) is known to be the *Riccati*-type ODE if dimensionless parameter $\zeta$ $\ll 1$ ($\zeta \to 0$), see Tab.1 below. Due to a very special character of the solutions of *Riccati*-type [10], there exists a possibility for them of sudden jumping of the meaning of solution at some meaning of time-parameter *t*.

In physical sense, it could be associated with the effect of *sudden gravitational slingshot* at the appropriate meaning of time-parameter *t* to change the trajectory of the small-body $m_3$ in the absolutely new unknown direction.

Thus, under assumptions of R3BP, we have derived that governing equations for the *quasi-circular* regular orbits of the moon (around the Planet in Solar system) could be reduced to the *Riccati*-type ODE above at first approximation. Such the equations should have the invariant form if the key governing (dimensionless) parameter $\eta$ is assumed to be in a range << 1 (even though it is not tending to 0).

Such an amazing fact let us evaluate the *forbidden zones* for Moons orbits in the inner Solar system or the zones of the meanings of distances $|\boldsymbol{R}_{2,3}|$ (*between Moon-Planet*) for which the motion of small body could be predicted to be unstable according to the basic features of the solutions of *Riccati*-type above.

It should not mean for all the cases of orbital *quasi-circular* motions to be under the effect of *sudden gravitational slingshot* at the appropriate meaning of time-parameter *t* (such a situation will depend on the initial data of the appropriate case). But we should note the existence of such a possibility for the safety of the future mission to Mars, for example.

## 2. The comparison of the forbidden zones for Moons orbits.

As we can see from Eq. (1.4), dimensionless parameter $\zeta$ appears to be the key parameter which determines the character of motion of the small 3-rd body $m_3$ (the Moon) relative to the 2-nd body $m_2$ (Planet). Let us assume such a dimensionless parameter $\zeta$ is varying (for *inner* Solar system) in a range from the meaning ~ $10^{-4}$ to

~ $10^{-2}$. In such a case, we could predict the *forbidden zones* for Moons orbits in the inner Solar system or the zones of the meanings of distance $|R_{2,3}|$ (*between Moon-Planet*):

$$\frac{|R_{2,3}|}{|R_{1,2}|} \cong \left(\zeta^{-1} \cdot \frac{m_2}{m_1}\right)^{\frac{1}{3}} << 1 \qquad (2.1)$$

Indeed, at given meaning of parameter $\zeta$ (for example, $\zeta = 10^{-3}$) we could evaluate it according to (1.4) as below (Tab.1):

Table 1. <u>Comparison of the forbidden zones for Moons orbits in the inner Solar system</u>

| Masses of the Planets (*Solar system*), kg | Ratio $m_2$ (Planet) to mass $m_1$ (Sun) | Distance $|R_{1,2}|$ (*between Sun-Planet*), AU | Ratio $m_3$ (Moon) to mass $m_2$ (Planet) | Parameter $\zeta \sim \left(\frac{m_2}{m_1} \cdot \frac{|R_{1,2}|^3}{|R_{2,3}|^3}\right)$ | Distance $|R_{2,3}|$ (*between Moon-Planet*) AU | Ratio $|R_{2,3}|/|R_{1,2}|$ |
|---|---|---|---|---|---|---|
| Mercury, $3.3 \cdot 10^{23}$ | $\left(\frac{0.055}{332,946}\right)$ | 0.387 AU | | $10^{-2} \div$ $\div 10^{-4}$ | 0.01 AU $\div$ 0.046 AU | 0.026 $\div$ 0.119 |
| Venus, $4.87 \cdot 10^{24}$ | $\left(\frac{0.815}{332,946}\right)$ | 0.723 AU | | $10^{-2} \div$ $\div 10^{-3}$ | 0.045 AU $\div$ 0.098 AU | 0.062 $\div$ 0.136 |
| Earth, $5.97 \cdot 10^{24}$ | 1 Earth = 332,946 kg | 1 AU = 149,500,000 km | **12,300**·$10^{-6}$ | $10^{-2} \div$ $\div 10^{-3}$ | 0.067 AU $\div$ 0.145 AU {Earth's Moon orbit ~0.003 AU} | 0.067 $\div$ 0.145 {Earth's Moon orbit ~0.003} |

| Mars, $6.42\cdot 10^{23}$ | $\left(\dfrac{0.107}{332{,}946}\right)$ | 1.524 AU | 1) Phobos $0.02\cdot 10^{-6}$ 2) Deimos $0.003\cdot 10^{-6}$ | $10^{-2} \eqsim$ $\eqsim 10^{-3}$ | 0.049 AU $\eqsim$ $\eqsim$ 0.104 AU {Phobos $6.3\cdot 10^{-5}$ AU = 9,380 km; Deimos $1.57\cdot 10^{-4}$ AU 23,460 km} | 0.032 $\eqsim$ $\eqsim$ 0.068 AU {Phobos $6.3\cdot 10^{-5}$ AU = 9,380 km; Deimos $1.57\cdot 10^{-4}$ AU 23,460 km} |
|---|---|---|---|---|---|---|

## 3. Discussion.

Let us discuss the results, presented in Tab.1. According to the expression (2.1), we have evaluated the *forbidden* distances |***R*** *2,3*| *between Moon-Planet* for unstable motion of small-bodies (such as moons of the Planets, or asteroids, or the spacecraft, drifting on the path around Planet) as below.

## 3.1. Mercury

The *forbidden* zone is located in the vicinity of 0.01 AU (the forbidden distance from small-body to the Planet). Besides, meaning 0.01 AU corresponds to the case of meaning of dimensionless parameter $\zeta$ equals to $10^{-2}$, so Eq. (1.4) should be transformed as below

$$\boldsymbol{R}_{2,3}'' + (1+0.01)\cdot\left(\frac{\gamma m_1}{|\boldsymbol{R}_{1,2}|^3}\right)\cdot \boldsymbol{R}_{2,3} = 0\,,$$

- but the meaning 0.046 AU could not be associated with the proper forbidden zone since condition $|\boldsymbol{R}_{2,3}|/|\boldsymbol{R}_{1,2}| \sim 0.119$ does not mean the case $|\boldsymbol{R}_{2,3}|/|\boldsymbol{R}_{1,2}| \ll 1$.

### 3.2. Venus

The *forbidden* zone is located in the vicinity of 0.045 AU (the forbidden distance from small-body to the Planet). Besides, meaning 0.045 AU corresponds to the case of meaning of dimensionless parameter ζ equals to $10^{-2}$, but ratio $|\boldsymbol{R}_{2,3}|/|\boldsymbol{R}_{1,2}| \sim 0.062$.

### 3.3. Earth

The *forbidden* zone is located in the vicinity of 0.067 AU (the forbidden distance from small-body to the Planet). Besides, meaning 0.067 AU corresponds to the case of meaning of dimensionless parameter ζ equals to $10^{-2}$, but ratio $|\boldsymbol{R}_{2,3}|/|\boldsymbol{R}_{1,2}| \sim 0.067$.

We should additionally note that the radius of the Earth's Moon orbit ~ 0.003 AU = 383,400 km < 0.067 AU (10,016,500 km).

### 3.4. Mars

The *forbidden* zone is located in the vicinity of 0.049 AU (the forbidden distance from small-body to the Planet). Besides, meaning 0.049 AU corresponds to the case of meaning of dimensionless parameter ζ equals to $10^{-2}$, but ratio $|\boldsymbol{R}_{2,3}|/|\boldsymbol{R}_{1,2}| \sim 0.032$.

We should additionally note that the radius of the martian moons Phobos $6.3\cdot 10^{-5}$ AU = 9,380 km $\ll$ 0.049 AU (7,325,500 km), but the radius for the Deimos $1.57\cdot 10^{-4}$ AU 23,460 km $\ll$ 0.049 AU (7,325,500 km).

The knowledge regarding the *forbidden* zone above (the forbidden distance from the trajectory of the small-body to the Planet) might be useful for martian mission in the nearest future.

**Conclusion.**

Previously, we have considered the equations of motion of three-body problem in a *Lagrange form* [1] (which means a consideration of relative motions of 3-bodies in regard to each other). Analyzing such a system of equations, we considered the case of small-body motion of negligible mass $m_3$ around the 2-nd of two giant-bodies $m_1$, $m_2$ (*which are rotating around their common centre of masses on Kepler's trajectories*), the mass of which is assumed to be less than the mass of central body.

In the current development, we have derived a key parameter η that determines the character of the motion of the small 3-rd body $m_3$ relative to the 2-nd body $m_2$ (Planet). Namely, by making several approximations in the equations of motion of the three-body problem, such the system has been reduced to the key governing *Riccati*-type ordinary differential equation.

Under assumptions of R3BP, we additionally note that *Riccati*-type ODE above should have the invariant form if the key governing (dimensionless) parameter η remains in the range $10^{-2} \div 10^{-3}$. Such an amazing fact let us evaluate the *forbidden zones* for Moons orbits in the inner Solar system or the zones of the meanings of distances $|\boldsymbol{R}_{2,3}|$ (*between Moon-Planet*) for which the motion of small body could be predicted to be unstable according to the basic features of the solutions of *Riccati*-type.

It should not mean for all the cases of orbital *quasi-circular* motions to be under the effect of *sudden gravitational slingshot* at the appropriate meaning of time-parameter *t* (such a situation will depend on the initial data of the appropriate case). But we should note the existence of such a possibility for the safety of the future mission to Mars, for example.

**Conflict of interest**

The author declares that there is no conflict of interests regarding the publication of this article.